\documentclass[conference, 10pt]{IEEEtran}
\usepackage{graphicx}
\usepackage{amsmath}
\usepackage{breakurl}
\usepackage{url}
\usepackage[nocompress]{cite}
\usepackage{verbatim}
\usepackage{stfloats}
\usepackage{subfigure}
\usepackage{tabularx}
\usepackage[protrusion=true, activate={true,nocompatibility}]{microtype}
\graphicspath{{figures/}}

\usepackage{color}

\newcommand{\ie}{{\em i.e.}}

\def\name{V-NDN }
\def\namenos{V-NDN}
\def\extname{Vehicular Named-Data Network }

\newcommand{\degree}{$^{\circ}$}

\begin{document}

\title{{\bf Vehicular Inter-Networking via Named Data \vspace{-0.3cm} }}

\author{
 \IEEEauthorblockN{Giulio Grassi\IEEEauthorrefmark{1}, Davide Pesavento\IEEEauthorrefmark{1}, Lucas Wang\IEEEauthorrefmark{1}, Giovanni Pau\IEEEauthorrefmark{1}\IEEEauthorrefmark{3}, Rama Vuyyuru\IEEEauthorrefmark{2}, Ryuji Wakikawa\IEEEauthorrefmark{2}, Lixia Zhang\IEEEauthorrefmark{1} \\ 
 }
 \IEEEauthorblockA{
   				 \IEEEauthorrefmark{1}Computer Science Department - University of California, Los Angeles, CA 90095, US\\
		}

\IEEEauthorblockA{
   				 \IEEEauthorrefmark{2}Toyota ITC - USA, San Jose, California.  \\
		}
\IEEEauthorblockA{
   				 \IEEEauthorrefmark{3} Universit\`{e} Pierre et Marie Curie (UPMC) - LIP6,  Sorbonne Universites - Paris, France. \\
   				e-mail: \{giovanni.pau\}@lip6.fr\\
		}
}

\maketitle


\begin{abstract}
In this paper we apply the Named Data Networking \cite{NDNWeb}, a newly proposed Internet architecture, to networking vehicles on the run. Our initial design, dubbed V-NDN, illustrates NDN's promising potential in providing a unifying architecture that enables networking among all computing devices independent from whether they are connected through wired infrastructure, ad hoc, or intermittent DTN.  This paper describes the prototype implementation of V-NDN and its preliminary performance assessment.

\end{abstract}
\section{Introduction}
\label{sec:intro}
In recently years research efforts in Vehicular Networking (VN) area have span into separate branches of networking research. Today's vehicles are connected through cellular networks and roadside nodes to centralized servers.  In addition, VN also expects to utilize results from ad hoc networking, since vehicles run pass each other unplanned in general, as well as from delay tolerant networking, not only because vehicles are often disconnected from time to time but also they can physically transport data from one location to another. 

To realize the vision where vehicles can communicate not only with the infrastructure but also with each other over any and all physical communication channels (including physically transporting data),
%
we take the Named Data Networking (NDN) as the starting point and develop a \textit{single} framework that enables vehicles to utilize any available channels to communicate in a completely infrastructure-free manner, both among themselves and with centralized servers to upload and retrieve data. We also built a prototype implementation.
Our design and implementation show a proof of concept and demonstrate the feasibility of integrating ad hoc, DTN, P2P functions all into one coherent network through a data-centric design that utilizes a shared application name space.

\section{A Sketch of Our Design}
\subsection{A Briefing on Named Data Networking}
\label{background}
In an IP network, node communicate by \emph{sending} IP packets to a specific destination address, which is obtained from DNS queries that translate an application level name to an IP address. In an NDN network, nodes communicate by fetching desired data.
NDN uses application data names for communication directly~\cite{VanSmThPlBriBra09-Networking}.
Since data names are defined by applications and exist independent from connectivity, this removes the need for IP address configuration (automatic or not), so that data exchange can happen as soon as \emph{any} physical connectivity comes into existence.

\subsection{\name for Mobile and Ad-hoc Networking}

\begin{figure}[t]
\begin{minipage}[h]{\columnwidth}
\includegraphics[width=\columnwidth]{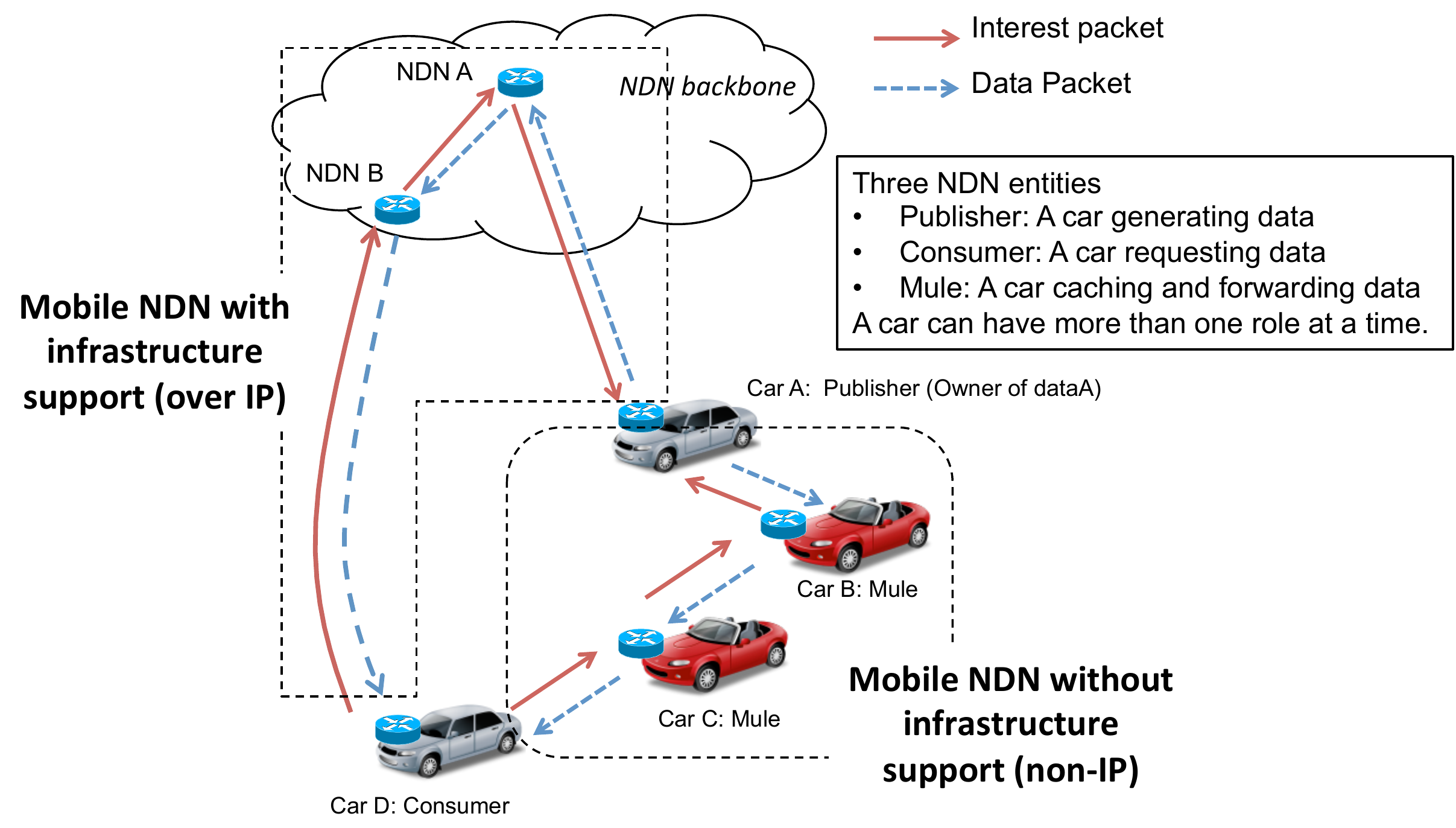}
\caption{\name \extname: supporting infrastructure and -less networks}
\label{fig:mobilendn}
\end{minipage}
\end{figure}

The description of NDN design so far \cite{VanSmThPlBriBra09-Networking, CCR-forward, Forwarding-TR} has mainly focused on wire-connected environment, with assumptions that do not fully match dynamic and ad-hoc mobile networking settings, such as a vehicular network.  We develop an instantiation of NDN for vehicle networking, \extname (\namenos), as described in this section.

In our system, a node (car) may be equipped with a variety of wireless interfaces such as 3G/ LTE, WiMAX, WiFi infrastructure or Ad-hoc mode, IEEE1901 (Power Line Communication), and 802.11p (DSRC/ WAVE).  Our design goal is to enable a car to utilize any of these interfaces to communicate with other vehicles and with infrastructure servers as needed by applications, as soon and as long as that interface is connected to other nodes. If more than one interface is available, one should be able to pick and choose the best one or use multiple in parallel.

NDN names the data to be fetched, the names used in communication are independent from which interface one wants to use, and from whichever nodes the data may come from.
Therefore \name can fully utilize any interface.  
Figure~\ref{fig:mobilendn} shows the elasticity of \name through different communication scenarios. 
The first scenario is when a car connects to infrastructure via 3G/ LTE, WiMAX, or WiFi as it changes its point of attachment while it moves.  In this case the car exchanges NDN interest and data packets with NDN routers located in wire-connected infrastructure.  We have installed NDN on WiFi routers which can communicate with cars directly over NDN; to communicate over cellular networks a car establishes IP tunnel to one of NDN routers in the Internet and send NDN packets over the IP tunnel (ex. Car-D tunnels an Interest packet to NDN-B over cellular channel). 

Another scenario is when a car exchanges NDN packets with neighbor cars locally through WiFi ad-hoc mode or/and 802.11p (DSRC/WAVE) in an entirely infrastructure-free manner.  For example, Car-D may send an Interest packet, which is received by the neighboring car (Car-C). The Interest packet can propagate hop-by-hop until it reaches the car with the requested data. In this case NDN packets are carried directly over the link-layer protocol. NDN extensions for local data circulation is explained in the next section.
As one can see from the figure, a car can communicate via different interfaces and send or receive packets through any of them (ex. Car-A and -D have connectivity to both cellular networks and WiFi).
%
The first NDN paper by Jacobson et. al. \cite{VanSmThPlBriBra09-Networking} introduced consumers, producers, and routers as three main types of entities in a wire-connected NDN network.
In contrast, in a \name network, a car may play any of the four roles \emph{simultaneously}: a data consumer, a data producer, a forwarder when it is connected to other vehicles, and a ``data mule'' when it physically carries data to other locations while having no connectivity to anyone else.

\subsection{\name for Local Data Circulation}
\begin{figure}[t]
\begin{minipage}[h]{\columnwidth}
\includegraphics[width=\columnwidth]{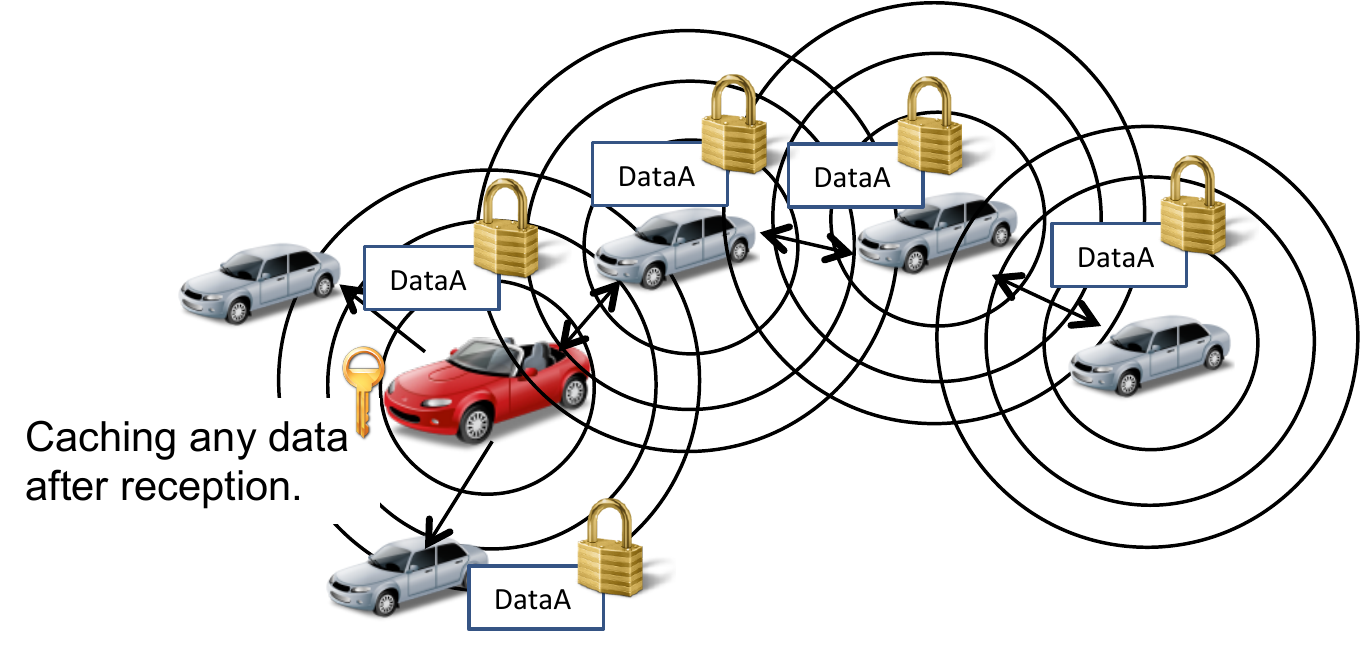}
\caption{\name: Packet are always singed and transmitted in broadcast; data is cached in the network at each node's content store.}
\label{fig:broadcasting}
\end{minipage}
\end{figure} 

To fit the ad hoc and intermittent characteristics of vehicle communication, as well as to take advantage of vehicle's resource properties (adequate storage capacity and power supply), the NDN node's implementation in \name must make necessary modifications from its wire-connected version.  
First, the high dynamics of a ``link'' among neighboring cars makes it infeasible to run a routing protocol to build the FIB. 
Second, since all communications are over wireless channel, one should take full advantage and cache overheard data packets (as opposed to only accept data which has a matching entry in PIT).
Third, either Interest or Data packets can be carried by running cars, even when they do not have wireless connectivity to anyone else.  Data can move out of the production location, by requests, or by car movements -- this makes it really important that we focus on reaching data, not geolocations.

To develop a traffic information dissemination application, instead of running a routing protocol, \name encodes geolocation information into data names, so that all Interest packets can be forwarded \emph{towards} the geolocation where the desired data is produced.  It is entirely possible, or even likely, that an Interest packet hits a car with the requested data, long before approaching the named location.

In a wired environment, an NDN node accepts a data packet only when it has an Pending Interest Table (PIT) entry for that data.  However, in high dynamic mobile scenarios, it may be infeasible to assume a stable chain of cars between the original consumer and where data is located to help return the data, a fundmental departure from wire-connected routers.  Thus in addition to use PIT in data forwarding whenever possible, \name also leverages wireless broadcasting nature and lets every node in the broadcasting range cache the received data regardless of whether it has a matching PIT entry or whether it needs the data itself.  Since \name treats mostly small data (e.g. traffic information) for M2M applications and not big files for video and music, and a car can also have relatively bigger cache space than mobile phones because of loose limitation of battery and space, this opportunistic caching strategy is considered advantageous in facilitating rapid data dissemination in highly dynamic environment.

In figure~\ref{fig:broadcasting}, Data-A is spread to neighboring cars and cached by all the receivers. Since these cars physically move around, they serve as data mules carry Data-A to wider area.  Having large number of mules enlarges data spreading areas, increases rendezvous opportunity of consumers looking for DataA with mules carrying DataA. 

In designing \name  we made an assumption that every node have accurate geolocation and all data names are linked to geolocation.  Contrary to fixed environment, it hardly maintains FIBs for every data because of publishers' movement. Therefore, geolocation is used to fetch data in \name.  A publisher does not advertise its data name to create FIB for the data as in the original NDN. 

Many vehicular applications fit the model where geographical locations can be used to identify the data object. The geolocation can be embedded in the naming. An application that requires parking availability on an city area, for example, could use a name that defines such area enabling \emph{any} vehicle in the area to respond.  Note that, although the data name embeds geolocation information, it is not naming the location but the data.  The geolocation information merely indicates where the requested data has been generated, thus it can be used as a hint as where to forward the Interest. 
In an earlier work we sketched out a basic name space design for a traffic information propagation application over NDN~\cite{NOMEN12}.  While car makers have traditionally focused on making vehicle body resilient in face of traffic accident, now a higher level of security measure is necessary when we start driving connected cars, to assure that vehicles do not fail because of a digital attack. NDN embeds security in its design by making each packet digitally signed. This provides the in-network support for information authenticity and trust mechanisms, and \cite{NOMEN12} also discussed solution directions to both protecting vehicle identities and preventing excessive false data from being injected into the system.
While many of the existing trust and security models~\cite{hu2006strong}\cite{laberteaux2008security}\cite{zhang2011survey}\cite{kamat2006identity} can take advantage of the infrastructure provided by NDN, the details are out of the scope of this work.

\section{Implementation}
\label{sec:implementation}

\begin{figure}[htb]
\begin{minipage}[htb]{\columnwidth}
\includegraphics[width=\columnwidth]{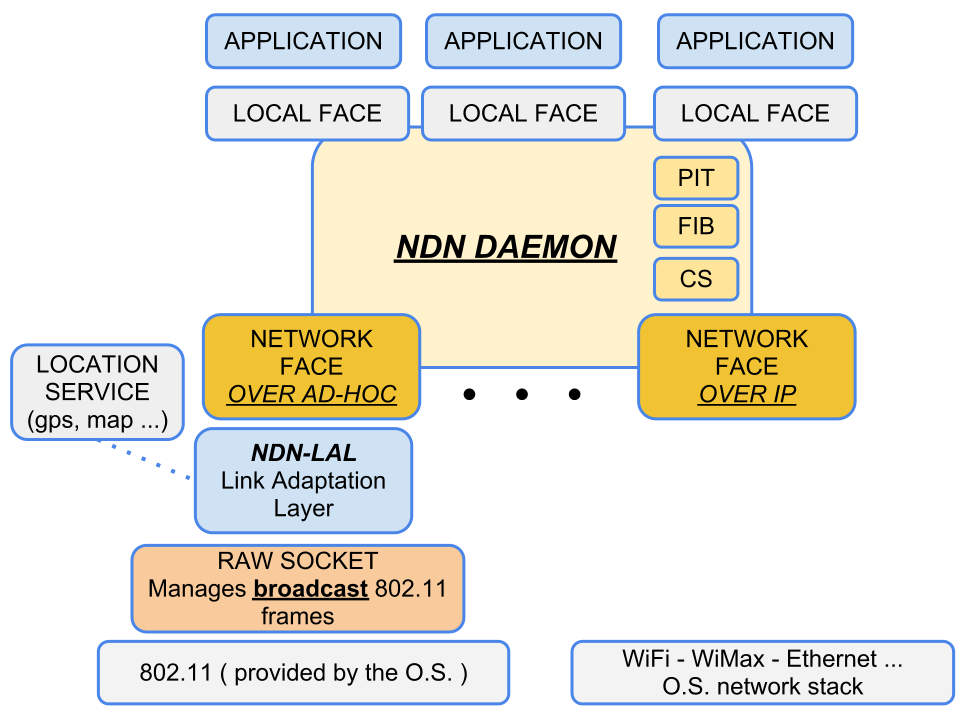}
\caption{\name logical architecture.}
\label{fig:architectureoverview}
\end{minipage}
\end{figure}

In the \name architecture a Vehicle can play 4 roles: data producer, consumer, forwarder (when connected), and data mule (during disconnection). Each vehicle may be multi-homed and the architecture is able seamlessly to forwards packets across different communication technologies. Finally, the design requires the vehicles to have a storage capabilities suitable to perform caching of data.  We designed and developed a \name prototype\footnote{The source code is available at https://github.com/named-data/vndn.}. The current version is developed under linux Ubuntu 12.04, however there are no kernel dependencies and the software is expected to run smoothly under any Linux distribution.  The architecture overview is depicted in figure \ref{fig:architectureoverview}. 

\begin{itemize}
\item {\bf NDN Daemon:} The NDN daemon provides the core named data networking capabilities.  It is the place for the key data structures such as the PIT, the FIB, and the Content Store. It's also where the name prefix-matching is performed and the forwarding decisions are taken. 
\item {\bf NDN Local Face:} The NDN Local Face provides an interface  between the Applications and the NDN daemon. In particular, the NDN Local Faces support  application name registration, Interest request by consumers and content delivery.  
\item {\bf NDN Network Face:} The NDN Network Face  provides the specific   adaptation functions coupled with the technology used in the communication. In vehicular case we use  IEEE802.11 based wireless technology in Ad Hoc mode for V2V, and  several  wireless technologies for the V2I case including WiMax, 3G, and WiFi based mesh networks.  For example for the 3G based technology the NDN Network Face provides the adaptation to the IP tunnel between the mobile node and the NDN node on the core. For the Wireless Ad Hoc case the NDN network face provides the interface with the \emph{Link Adaptation Layer} which provides wifi broadcast support for the vehicular environment.  
\item {\bf Link Adaptation Layer:} The \emph{Link  Adaptation Layer} (LAL) is conceptually a  2.5 layer designed to enable NDN to efficiently take advantage of specific technologies. In the vehicle-to-vehicle scenario which is usually performed using either WiFi or DSRC broadcast transmissions would greatly benefit NDN; however IEEE802.11 broadcast support is practically inexistent thus a number of tasks are surrogated by the LAL~\cite{IEEE80211}\cite{IEEE80211p}. In particular, the acknowledgment/retransmission mechanisms and the spatial awareness of the packet progress. Finally, in order to minimize overhead, the link adaptation layer sends all the packets as L2 broadcast, the NDN packet format is directly overlaid over the wifi frame, no other L3 protocol is used thus implementing a native NDN network layer.  
\item {\bf Location Services:} The location services provides the support to the applications as well as to the link  adaptation layer. In particular the location services provide reverse-geocoding capabilities, as well as high level functions on distance and heading. The link adaptation layer uses the location services to geographically scope the communication. 
\item {\bf Cache:} Each node is equipped with a cache that stores contents that go through each node. In the current implementation the cache items have no expiration and virtually infinite. \footnote{The current cache is stored in main memory and is sized to 10GB, the next version will also employ a disk-based cache (content store).}
\end{itemize}

\begin{figure*}[t!h]
\vspace{-0.2in}
\begin{minipage}[t!h]{\textwidth}
\centering{
\subfigure[\name infrastructure assisted V2V: packet route]{
\includegraphics[width=0.30\columnwidth]{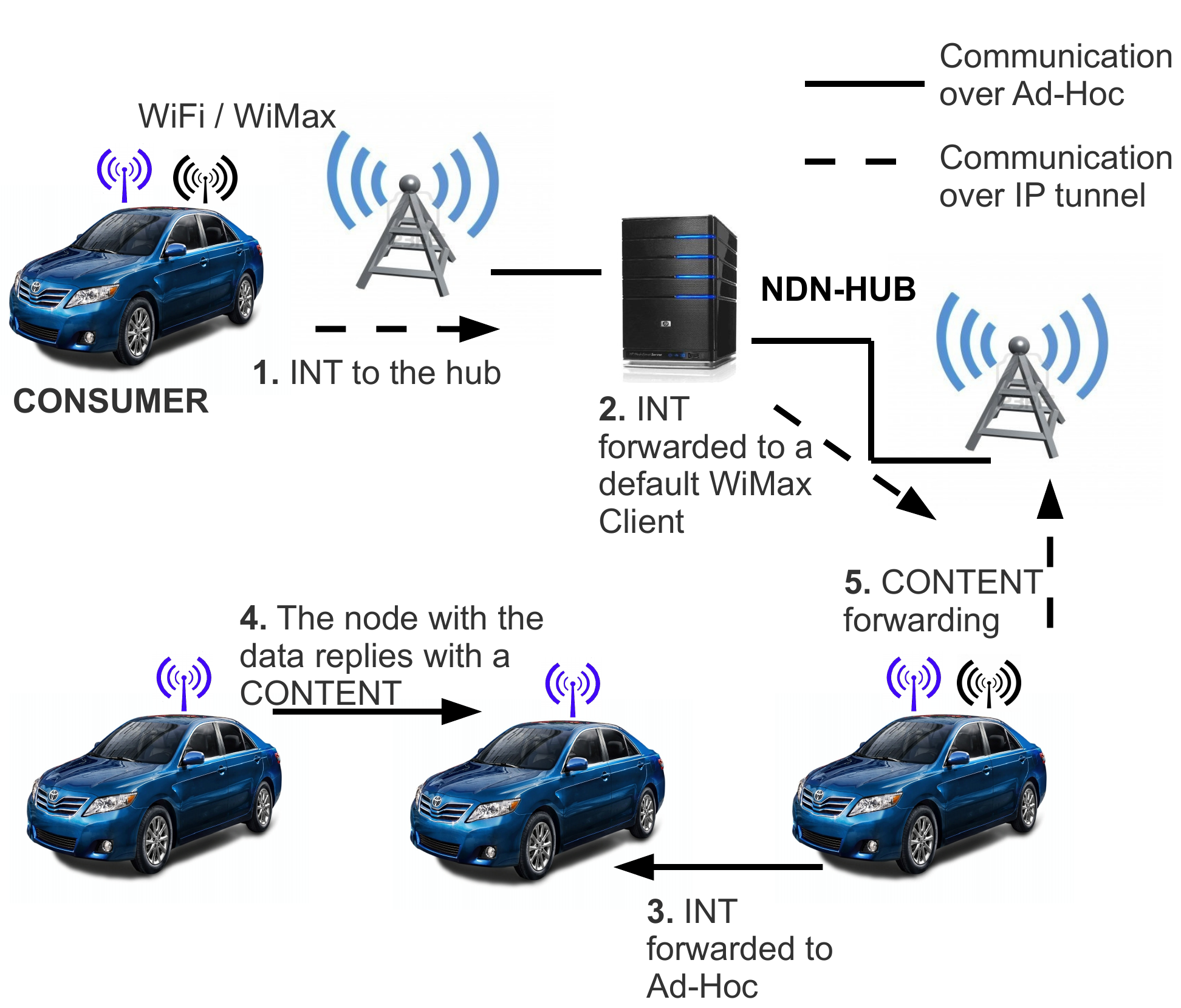}
\label{fig:V2VRoute}
}
\hfil
\subfigure[\name I2V packet route]{
\includegraphics[width=0.30\columnwidth]{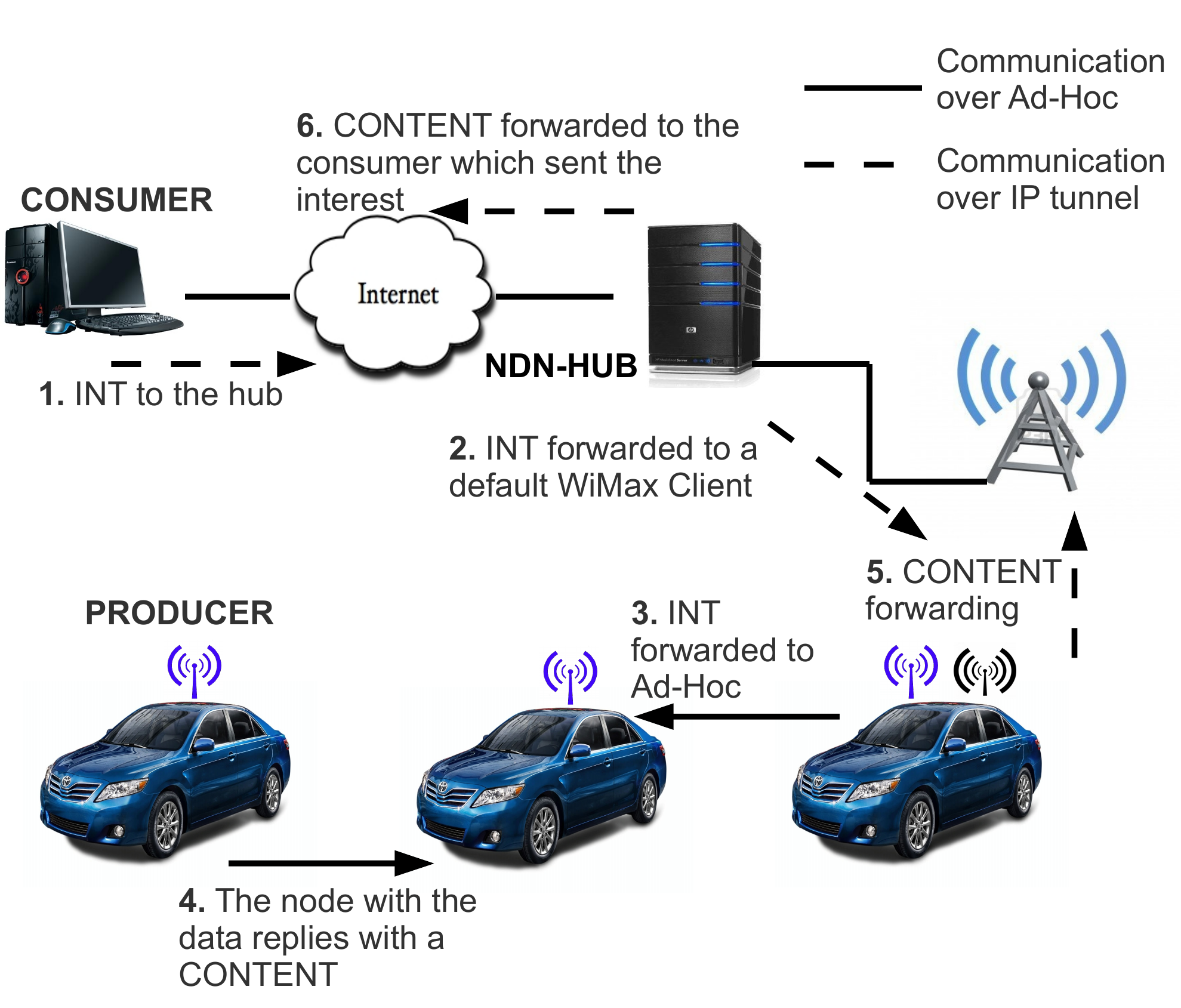}
\label{fig:I2VRoute}
}
\hfill
\subfigure[\name V2I: packet route]{
\includegraphics[width=0.30\columnwidth]{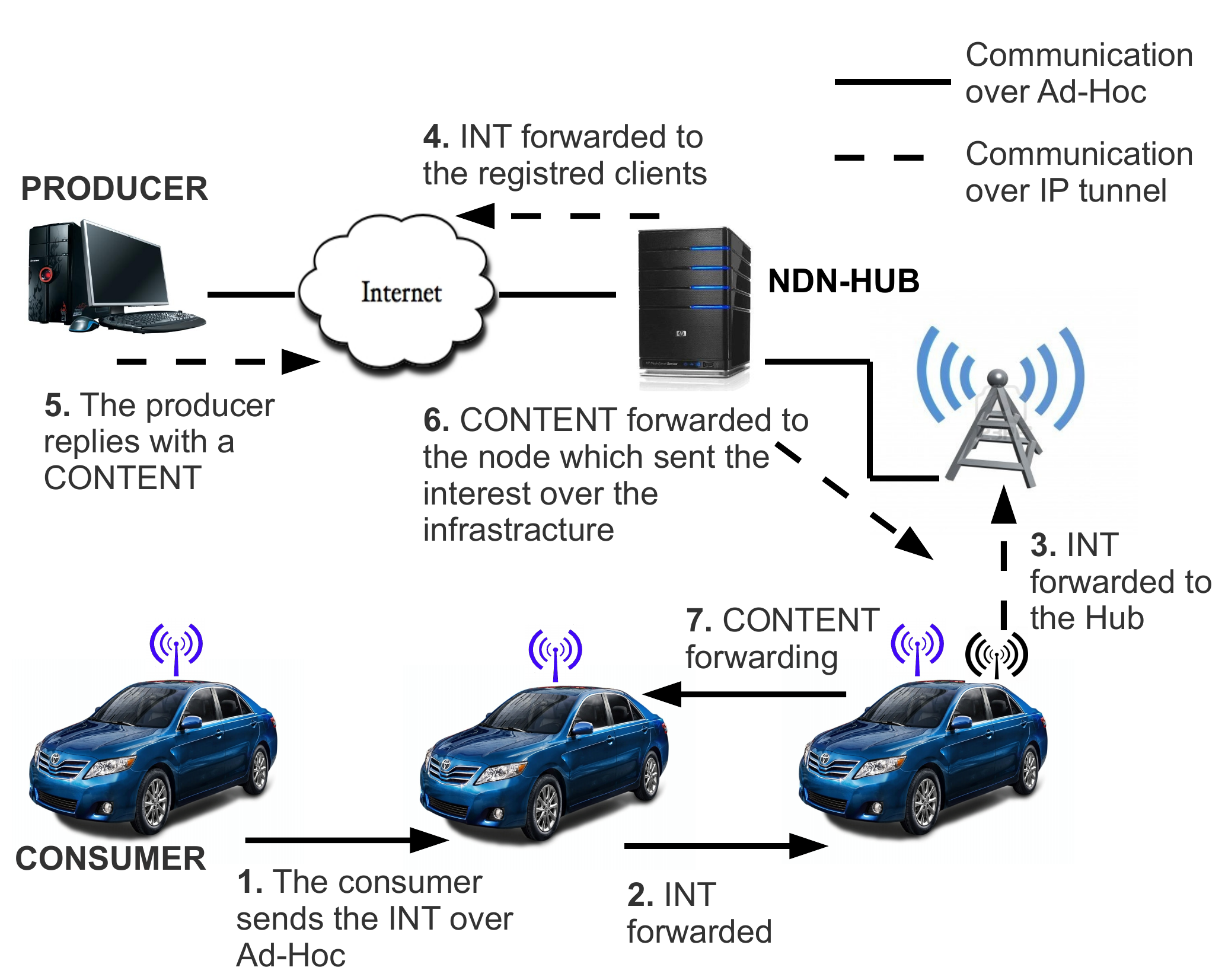}
\label{fig:V2IRoute}
}  
}
\caption{Packet route in \name: V2I, I2V, and infrastructure assisted V2V scenarios}
\label{fig:packet-paths}
\end{minipage}
\end{figure*}

\subsection{Enhancing WiFi Broadcast for V2V communications}
\label{sec:wifibroadcast}
We use L2 WiFi broadcast for all the V2V communications. However current IEEE802.11 standards provide a practically inexistent broadcast support. In particular, broadcast transmissions are not protected  (\ie  No RTS/CTS), and there are no acknowledgment/ retransmission mechanisms for broadcast traffic in the standards~\cite{IEEE80211} thus broadcast communications in WiFi suffer high loss rates.  This is exacerbated by the nature of vehicular networks that feature relatively short link durations, dynamic partitioning, and fast changing topologies~\cite{rowstron2009characteristics}. In contrast, \name interest forwarding and content retrieval greatly benefits from efficient and resilient broadcast transmissions as it enables opportunistic forwarding, delivery and caching.  

We designed and implemented a simple set of mechanisms to improve broadcast support in WiFi-based VANET communications; those mechanisms take place in the Link Adaptation Layer.  In our architecture the LAL determines if a packet is making progress (1-hop progress). It performs number of typical link layer tasks such as for example \emph{handling acknowledgments}, \emph{detecting loss/disruption}, and \emph{managing retransmissions}. \name uses an implicit acknowledgment system designed for vehicular communications and extends the  work  in \cite{wang2012rapid}. Each vehicle is assumed to have a GPS and Location service based on a digital map\footnote{In our prototype, we implemented a tiger-map based location service which supports reverse geocoding, distance computation, and several other spatial awareness services.}. The link adaptation layer  is able to take advantage of the location services to provide \emph{spatial awareness} and scope the forwarding and retransmission policies according to the actual situation on the ground. Ideally the link adaptation layer aims at a greedily forwarding each packet at the maximum distance from the current transmitting node thus resulting in a faster packet progressing. The ideal next hop is therefore the furthest reachable vehicle in each of the selected forwarding directions. The main task of the LAL is to ensure the packet is making progress and, possibly, making progress at the maximum speed. 

The implicit acknowledgment mechanism is implemented at the LAL and performed every time a packet is received from the NDN Network Face. It relies on two different timers to achieve the maximum forwarding distance. First small random timer used to randomize the transmission time and hence reduce the collision probability between equally-probable nodes.  Second a  timer designed to rank vehicles inversely to their distance from the last-hop  and defined as $\frac{1}{D(sender,receiver)}$, where $D$ distance computed using the location service.  A packet is considered acknowledged by a node if it  over-ears the same packet broadcasted  by a car other than the last forwarder. In urban scenarios, a finer spatial awareness may be introduced at road intersections. A packet may be considered fully acknowledged if a retransmission is eared from each of the roads stemming from the such intersection. This approach allows the NDN level and/or the applications developers to apply different policies according to the application requirements. 

\section{Demonstration Setup}
\label{sec:demonstration}
\begin{figure*}[htdp]
\begin{minipage}[htdp]{\textwidth}
\centering{
\subfigure[\name Driving Route, Clock Wise  one block around Parking Lot 8]{
\includegraphics[width=0.40\columnwidth]{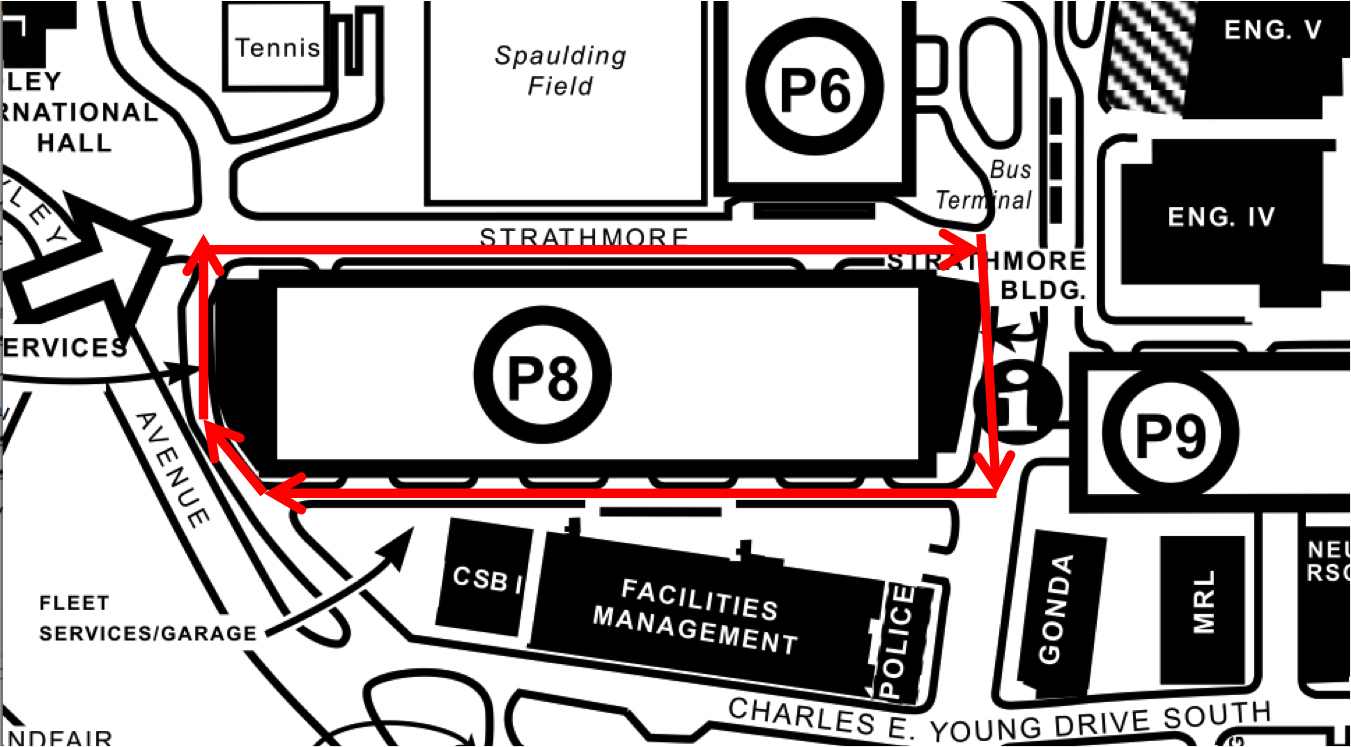}
\label{fig:clockwiseroute}
}
\hfil
\subfigure[\name Driving Route, Clock Wise two blocks around Parking Lot 8]{
\includegraphics[width=0.40\columnwidth]{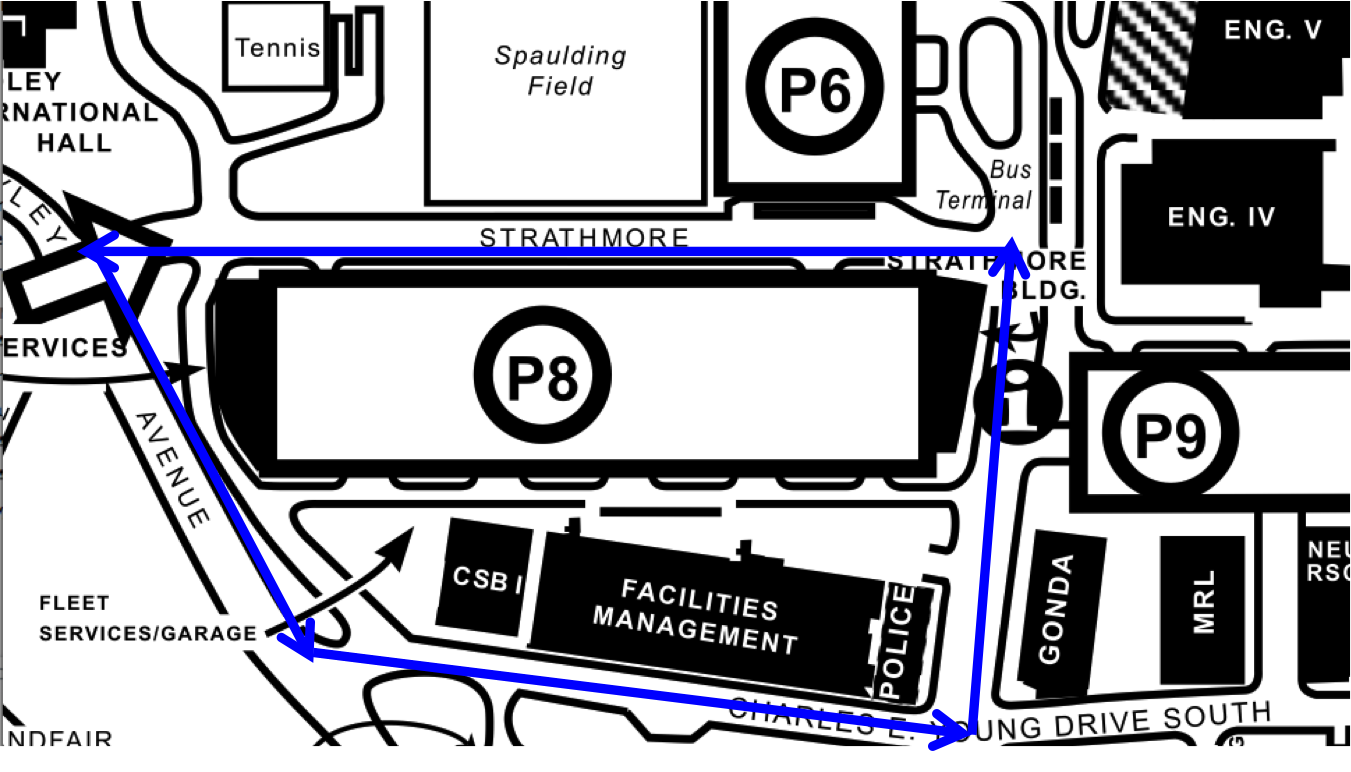}
\label{fig:counterclockwiseroute}
}
}
\caption{Mobility Patterns}
\label{fig:MobilityPatterns}
\end{minipage}
\end{figure*}

\name has been implemented as proof of concept prototype at UCLA and tested in campus  using  UCLA  Vehicular Testbed. Several experiments were performed on November 25, November 30, and December 1st, 2012, the first day was a clear day while the other two were raining days. The experiments involved multiple vehicles. In particular we used 6 vehicles on Nov 25, 2012, 10 vehicles on November 30th 2012, and 9 vehicles on December 1st 2012.  We designed and implemented two NDN applications for the vehicular domain; the {\bf Info-Traffic} application and the {\bf Road-Photo Application}. The info-traffic application emulates the request of traffic information for a specific area and the response from a vehicle that has recently been in such area likewise the road-photo application emulates the request of a picture for a specific area and the response from a vehicle that has recently been in that area.  The  area is encoded in the name and rather than coordinates we refer to intersections and streets stemming from that intersections \ie /traffic/westwood-at-strathmore/ would refer to the traffic information from the area close to the Westwood-Strathmore intersection in the UCLA campus.  A car that has been or is currently in such location will respond to the interest reporting the proper traffic information. Is worth noticing that multiple responses may be generated by one interest  however, in the current implementation,  only the first response is considered.  We employed two different types of mobility pattern: \emph{Urban Platooning} with the cars going around a large urban block or a parking lot in a single line platoon, and a more realistic mobility which we call \emph{double-clock} where a part of the available vehicles run in clockwise loops around a block and another disjoint set of cars runs in a counter-clockwise block in an adjacent block thus vehicles from different loops can communicate roughly only half of the time; figure \ref{fig:clockwiseroute} and \ref{fig:counterclockwiseroute} show the route details for the two groups of vehicles.  

The experiments were designed to investigate \name behavior in  the following communication scenarios typical for the vehicular applications domain.
\begin{itemize}

\item {\bf V2V}: Interests and content travel through WiFi Vehicle-to-Vehicle Connections thus the content addressed is relatively local (\ie few hops away). Applications such as assisted left-turns and accident warnings may benefit from this type of communications (figure \ref{fig:V2VRoute})

\item {\bf I2V}: Lets consider a traffic monitoring system that constantly requests the current status for various critical intersections. The consumer node resides on the wired network (\ie traffic control center) and the producer node resides on the vehicular network \ie vehicles that are close or have been recently close to a point of interest (figure\ref{fig:I2VRoute}).

\item {\bf V2I}: The producer is connected to the wired network while the consumer is on the vehicle.  A typical example is when a vehicle wants to retrieve data about the traffic in the destination and direct v2v routing may be inefficient or even not possible (figure\ref{fig:V2IRoute}).
 
\item {\bf resilience to disruption}: Vehicular networks are prone to disruption, the link duration is relatively short and the topology is continuosly changing resulting in dynamic network partitions \cite{rowstron2009characteristics}. We designed a set of experiments to reproduce this phenomena and study NDN behavior/performance  in such harsh conditions. 

\item {\bf in-network storage}: One of the great innovations of the Named Data Network  paradigm is to introduce caching in the network.  This feature improves efficiency and  turns out to be  essential for VANETs.  
\end{itemize}

\subsection{Field Experiments}
We performed a number of field experiments for both applications varying the mobility patterns and the type of communication involved (\ie V2V, V2I/I2V, V2V2I). Each experiment was repeated several times. The experiments are summarized in table \ref{tab:experiments}
\begin{table}[htdp]
\caption{Experiment Summary}
\begin{center}
\begin{tabular}{|c|c|c|c|}
\hline
Exp. \#&Application&Type & Mobility\\
\hline
A.1&Info-Traffic&V2V&P8:Still \& \\
&&&Urban Platoon\\
\hline
A.1&Info-Traffic&V2V&P8:Still \& \\
&&&Urban Platoon\\
\hline
A.3&Info-Traffic&V2V&Fig.~\ref{fig:clockwiseroute}~\&~\ref{fig:counterclockwiseroute}\\
\hline
A.4&Road-Photo&V2V&Fig.~\ref{fig:clockwiseroute}~\&~\ref{fig:counterclockwiseroute}\\
\hline
\hline
B.1&Info-Traffic&V2V&P8:Still \& \\
&&&Urban Platoon\\
\hline
B.2&Info-Traffic&V2V&P8:Still \& \\
&&&Urban Platoon\\
\hline
B.3&Info-Traffic&V2V2I&Fig.~\ref{fig:clockwiseroute}~\&~\ref{fig:counterclockwiseroute}\\
\hline
B.4&Road-Photo&V2V2I&Fig.~\ref{fig:clockwiseroute}~\&~\ref{fig:counterclockwiseroute}\\
\hline
\end{tabular}
\end{center}
\label{tab:experiments}
\end{table}%
\subsubsection{{\bf Experiments A1, A2, B1, B2}}  are preliminary experiments performed on the rooftop of UCLA parking structure  P8 with \emph{no mobility} or very basic \emph{slow urban platooning}. UCLA P8 rooftop is 265m long and 76m large. The wireless range was  to cover the whole roof at the operating frequency of 2.4GHz; however due to a very large number of competing transmitters (net-stumbler reports about 100 access points that can be eared in that area) the radio range using broadcast messages unexpectedly ranged from 1m to 30m. Additional experiments performed in ideal conditions at the \emph{El Mirage Dry Lake Bed} ( 34\degree 37'50.10"N, 117\degree 34'14.64"W), in California show that with no-interference an 2.4GHz wifi interface can achieve up to 550m.

\subsubsection{{\bf Experiments A3, A4, B3, B4}} were performed on the roads around UCLA structure P8. The pool of 10 vehicles has been divided in two groups of 6 and 4 cars respectively. The smallest group of vehicle was running clockwise around the P8 Block as shown in fig. \ref{fig:clockwiseroute}. The second (and largest) part of vehicles was running counterclockwise to cover a larger road block which also includes the UCLA P8 structure see figure \ref{fig:counterclockwiseroute}. Car speeds ranged from 6.3$_{m/s}$ to 21.2$_{m/s}$  \ie {\raise.17ex\hbox{$\scriptstyle\sim$}}14 to 47mph. The mobility pattern allows vehicles traveling in opposite direction to meet however prevents a continuos connectivity between the two partitions. In particular, the small loop shown in figure \ref{fig:clockwiseroute} has a total length of  827 meters; while the larger loop shown in figure  \ref{fig:counterclockwiseroute} has a total length of 1,023 meters, the common areas amount to {\raise.17ex\hbox{$\scriptstyle\sim$}}514 meters, thus the vehicles in the larger loop have a chance to meet the vehicle in the shortest loop about half of the time. It is important to notice that having shared segments on the road is not a sufficient conditions for vehicles to meet and exchange data. Traffic lights, road traffic, and other urban artifact play an important role in determining wether or not two vehicles will meet. Specifically, there are 4 traffic lights around the small loop  and 6 traffic lights on the large loop. The experiments are conducted from 11am to 8pm and they factor rush- and regular hour traffic as well as pedestrian traffic at the intersections. Traffic lights and pedestrian traffic created dynamic partitioning and  voids during the experiment as it happens during the regular course of traffic in urban scenarios~\cite{rowstron2009characteristics}. While in motto, and therefore away from the roof of the parking lot, we observed a longer  wifi range  up to 318 meters.

\subsubsection{{\bf Hardware/Software Setup}} \name was installed in low-cost netbooks the \emph{Asus EeePc 1011CX} powered by an Intel Atom N80 at 1.6GHz; each node had 2GB of ram and 160GB of hard drive. Each node was retro-fitted by us with a MIMO-capable WiFi Card the \emph{Unex DNXA-92} with a specifically customized firmware to allow world-wide operations and IBSS at 5.8GHz.  The Unex DNXA-92 WiFi card is based on the Atheros  AR9280 which is supported by the Linux standard driver Ath9k. The maximum transmission power at the antenna connector is declared by the manufacturer to be  27dBm for 2.4GHz and 30dBm for the 5.8Ghz band\footnote{ During the demo only 6 nodes had the Unex DNXA-92 while the others had a standard DNXA116 thus the results factor different NICs in  hardware platform. We are planning a future test using only DNXA-92}. We used dual band omni-directional antennas the Hyperlink HG2458-7RDR-NM featuring a gain of 4.5dBi at in the 2,400-2,500MHz band and 7dBi in the 5,150-5,850MHz band, the antenna has circular polarization. All the experiments have been performed using  the 2.4GHz band (Channel 1).  Two nodes out of 10 were equipped with a second WiFi interface operating in Infrastructure mode. We used an Ubiquity Networks SR71USB which is also supported by the Linux drivers (carl9170). Furthermore two other nodes were equipped with a WiMax usb based interface connected to the UCLA-GENI WiMax testbed and operating at 2,690MHz, specifically we used the Teltonika UM62x1 which supports multiple antennas and spatial diversity. We also operate directly the WiMax base station the NEC INW-INTFC and at the time of experiments no other users were connected.  We used  an off-the-shelf Ubuntu Desktop 12.04 as operating system and \name was installed in the user space, no kernel hacks are required.  \name transmits broadcast packets in the WiFi Interfaces operating in Ad Hoc mode while tunnels the NDN traffic over IP for the other interfaces (see figure \ref{fig:packet-paths}). A \name hub is connected at the other tunnel endpoint and prefers the appropriate forwarding tasks.  
The experiments were controlled by an additional 4G link provided by Clear Wireless, however such link was used only for experiment control and telemetry purposes and not as experimental link. 

\subsection{Preliminary Results}
We used rsyslog to collect the header of all the packets through the network and to log the software major events (\ie forwarding, routing, etc), this resulted in several hundred megabytes of log data helpful to gather insights and understand the capabilities of NDN. 

\begin{figure*}[htdp]
\begin{minipage}[htdp]{\textwidth}
\centering{
\subfigure[\name Info-Traffic application CDF number of Link Adaptation Layer retransmission]{
\includegraphics[width=0.47\columnwidth]{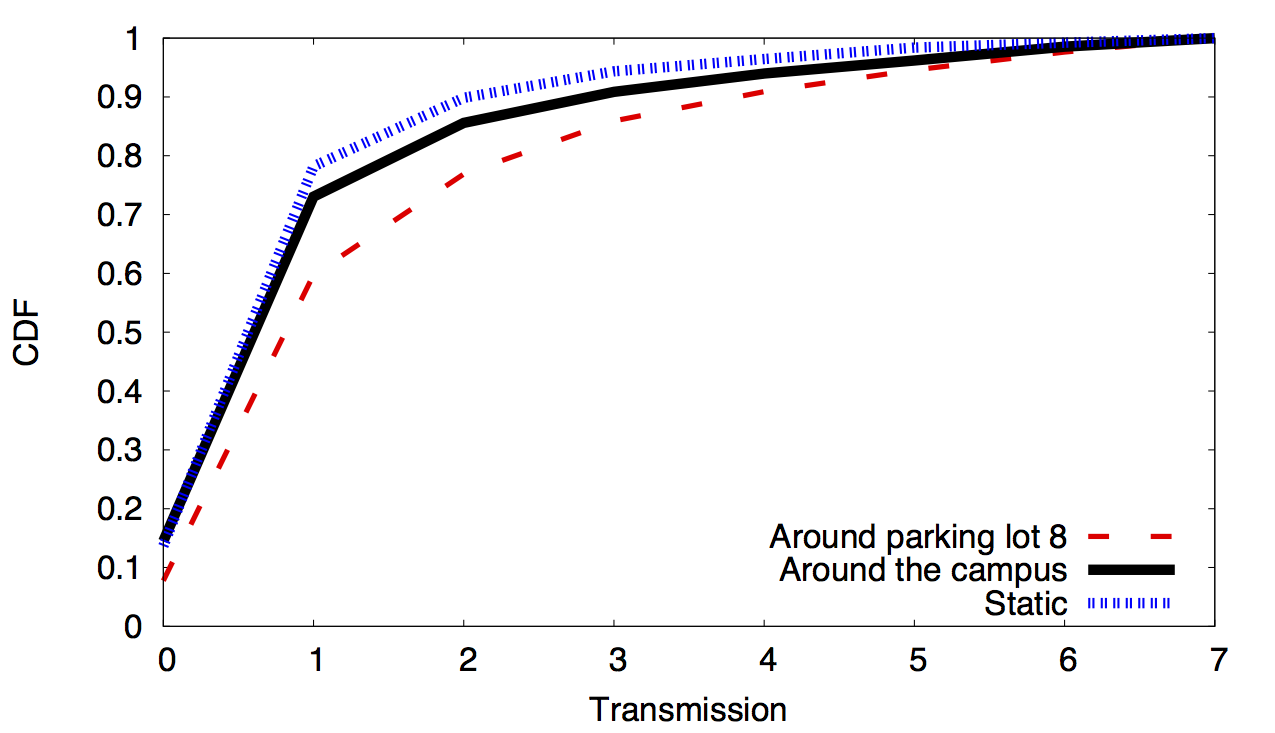}
\label{fig:trafficappcdfretransmissions}
}
\hfil
\subfigure[\name Info-Traffic application: CDF Response time  -  from the first Interest to the corresponding data ]{
\includegraphics[width=0.47\columnwidth]{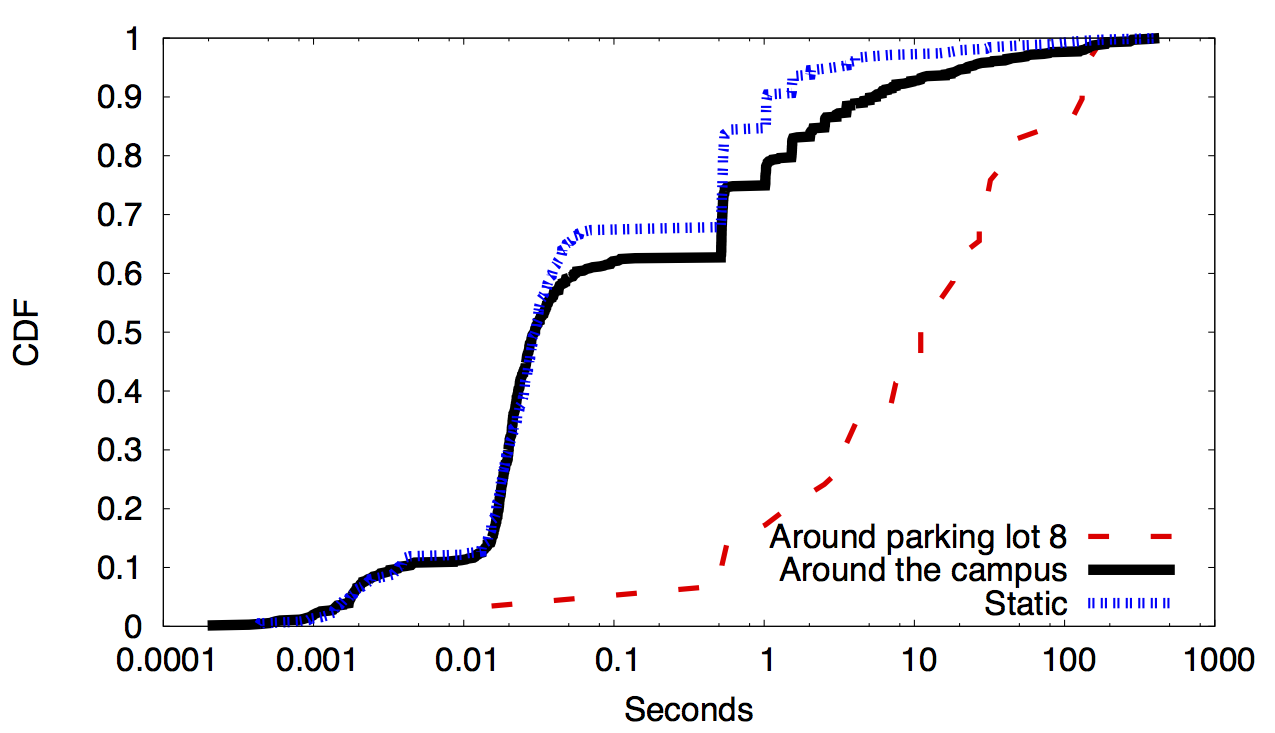}
\label{fig:trafficappcdfresponsetime}
}
}
\caption{RTT Analysis}
\label{fig:preliminaryresults}
\end{minipage}
\end{figure*}

\subsubsection{{\bf V2V experiments }} 
Those experiments were mainly performed on November 30th and December 1st 2012, vehicles had 1 wifi interface in Ad Hoc mode. In order to support NDN using broadcast we devised mechanism for implicit acknowledgment and retransmissions that allows interest and data to make progress as detailed in section \ref{sec:wifibroadcast} the mechanism is implemented in the Link Adaptation Layer which conceptually is at layer 2.5.  \name Link Adaptation Layer manages link layer retransmissions and acknowledgments. In particular, the proposed approach tries limit the rebroadcasting of packets choosing as potential relay the nodes that are the furthest from the current one. \name LAL has embedded spatial awareness using digital maps. We designed a mechanism to allow applications to chose the degree of spatial acknowledgment; for instance at each intersection an application may require packets to be acknowledged by any or all  the road directions. The digital map and the required functionalities are implemented by the mean of a system-wide  stand-alone location service. The acknowledgment mechanism is implicit \ie the node over ears the same packet forwarded by somebody else; if no acknowledgement is detected \name link adaptation layer retransmits the packet up to \emph{max-retransmission} which was set to 7 throughout the experiments. We studied several factors:
\begin{itemize}
\item \emph{(i)} The number of retransmission needed to forward a packet by 1 hop using wifi broadcast 
\item \emph{(ii)} the response time at the application level, 
\item \emph{(iii)} the use of cache vs interest forwarding (both in regular nodes and relay-only nodes), 
\end{itemize}

Figure \ref{fig:trafficappcdfretransmissions} shows the CDF for the \name Link Adaptation Layer transmissions for the info-traffic application in different experimental scenarios (Table \ref{tab:experiments} experiments A1, A3). The chart shows that for the static case about 75\% of the packets need just one LAL retransmission. In the mobility scenario this figure goes down to about 65\% however the type of mobility (either on the P8 roof or on the roads)  has a negligible impact on the retransmissions.  It is interesting to notice that roughly 95\% of the packets are acknowledged in 5 retransmissions or less  and therefore the max-retransmission currently set to 7 is clearly an overshoot. Finally, 15\% of packets (for the static case) were acknowledged to the \name LAL before being actually transmitted and therefore they never went on the air.

The response time cumulative distribution function for the Info-Traffic application (Table \ref{tab:experiments} experiments A1, A3), is shown in figure \ref{fig:trafficappcdfresponsetime}. The chart shows that, in the static case,  about 75\% of the info-traffic interests got a response in less than 1 second. For the mobility case is interesting to note that experiments performed looping around on the roof of parking lot 8 performed worse than the experiments on the roads around campus. We believe this is due to the effect of traffic lights which introduce aggregation points and relatively long breaks in the mobility flow thus facilitating connectivity. In addition, the UCLA parking structure P8 is located in a midst of wifi access points and therefore has a very high noise background.

\begin{table}[htdp]
\caption{Cache vs Forwarding among all nodes}
\begin{center}
\begin{tabular}{|c|c|c|c|}
\hline
& Static & Around P8 & Around \\
& & & Campus\\
\hline
Use of Cache&2814&5280&3124\\
\hline
Interest Forward&4834&7068&17784\\
\hline
\end{tabular}
\end{center}
\label{tab:cachevsforwardinggeneral}
\end{table}%
Table \ref{tab:cachevsforwardinggeneral} shows for each mobility scenario the role of caching considering all the nodes in the network. In particular chatting contents appears to be more important in mobility. This is expected as the topology changes relatively fast.  
\begin{table}[htdp]
\caption{Cache vs Forwarding among Consumers and Mules}
\begin{center}
\begin{tabular}{|c|c|c|c|}
\hline
& Static & Around P8 & Around \\
& & & Campus\\
\hline
Use of Cache&96&3054&1840\\
\hline
Interest Forward&715&4685&13195\\
\hline
\end{tabular}
\end{center}
\label{tab:cachevsforwardingconsumersmules}
\end{table}%
In the aim to better understand the role of in-network caches for \name we  excluded from the dataset producers and analyzed the cache/forward statistics for consumer and mules. Results are shown in table  \ref{tab:cachevsforwardingconsumersmules}; the data confirms that caches are more beneficial during mobility and particularly when mobility happens in a relatively restricted area (rooftop of parking structure P8).
Observing  mule-nodes  (\ie no interests or content are sourced in this nodes) only, the role of in-network caches becomes even more prominent;  in this case about 66\% of the traffic is found in the local cached and therefore the interest is not forwarded further. 

A second reference application was tested during the experimental run, a consumer issues an interest for a picture of a certain area, the vehicle closer to that area takes a picture and delivers it back to the consumer. We implemented the application taking realtime pictures. The consumer issues an interest with a name that sounds "/picture/westwood/strathmore/" indicating it is requesting a picture from a camera that is on board of a vehicles close to the Westwood-Strathmore intersection. A vehicle in that area, if any, will turn on the camera, take a snapshot, and send it back.  While our design is far from optimal and does not employ any optimized segmentation or coding technique we were able to retrieve  51 camera shoots. Photos had an average size of 6.3KB resulting on average in 5 chunks; in order to consider a photo received all the chunks need to arrive at the consumer.  The road-photo application experienced a response time  averaging in 81 seconds for the mobile case (fig. \ref{fig:MobilityPatterns}), and 28 seconds for the static case; the median point was 55.6 seconds, and 1 second respectively. The sparse nature of our network has affected the response time for the road-photo application which suffers from disruption much more than the info-traffic application. The content for the info-traffic application is contained in one single content packet while the road-photo application requires several packets to arrive at the consumer. We believe that applying optimization techniques such as Network Coding and a more sophisticated file encoding the results can improve by far. 

\subsubsection{{\bf The role of the Infrastructure and the V2X Scenario}}  

\begin{figure}[htb]
\begin{minipage}[htb]{\columnwidth}
\centering{
\includegraphics[width=0.5\columnwidth]{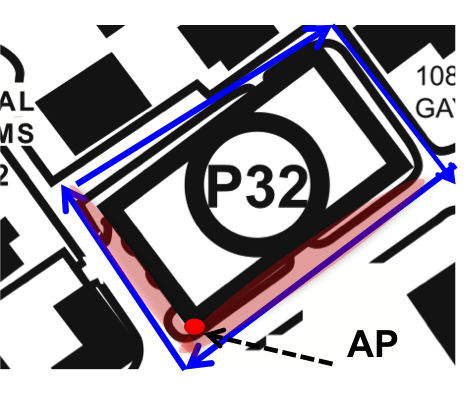}
\caption{\name V2X experiment: Mobility pattern and AP placement.}
\label{fig:car2XLot32}
}
\end{minipage}
\end{figure}

\begin{figure}[htb]
\begin{minipage}[htb]{\columnwidth}
\centering{
\includegraphics[width=1\columnwidth]{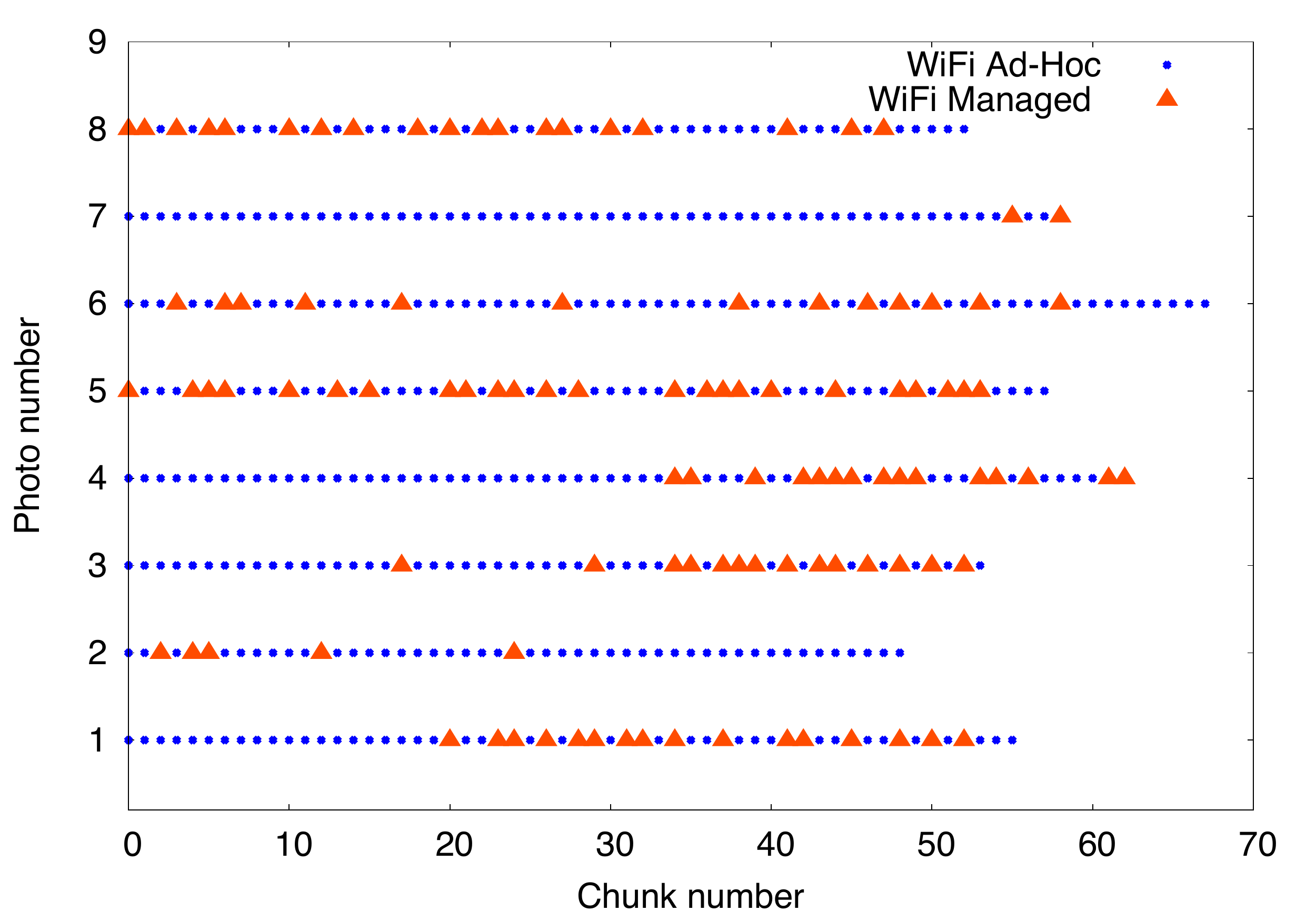}
\caption{\name V2X experiment: Link to Content allocation, each dot is a picture's chunk. }
\label{fig:V2Xlin2content}
}
\end{minipage}
\end{figure}


In the November 25th and November 30th experiments we had few multi-homed vehicles. In particular, on November 30th two nodes were multhomed and equipped with both WiFi (V2V) and WiMax interfaces while on November 25th two nodes had  WiMax interfaces and one node had an extra WiFi interface in Infrastructure mode on top of the regular V2V interface.  The results are mainly qualitative and they are to be used mainly as a proof of concept rather than a performance evaluation. 

On November 30th 2012, 25 contents were received from the WiMax interface in one of the two multi-homed nodes and then delivered to the destination, those contents were produced by a \name node sitting on the wired network behind the WiMax ASN-Gateway.  On the november 30th experiments we were able to get 388 contents from the WiMax and 34 contents from the WiFi Infrastructure. 

In a second experiment  designed to better understand this phenomena we used two vehicles running around parking lot 32 in a clockwise fashion. On one corner of the parking structure we setup an access point and connected it to the campus infrastructure (fig. \ref{fig:car2XLot32}).  The \emph{consumer}  was equipped with 2 WiFi interfaces, one operating in  Ad-Hoc mode and a second one operating in Infrastructure mode. The \emph{producer}  was equipped with 1 WiFi interface configured in Ad Hoc mode and 1 WiMax interface. In this experiment we considered a photo fetching scenario; the consumer requests a photo to be taken by the producer at regular intervals. Interest and data packets both traveled to any interface that is available at the time of transmission. Photos were taken in realtime upon receiving the interest via the computer on-board camera, their sized between 68KB and 100KB. Each photo is split in several chunks of 1300 bytes. We accounted how on which interface chunks are transmitted as shown in figure \ref{fig:V2Xlin2content}

\subsubsection{{\bf Data survives the Producer}}  
Once a content has been spread  on the network, it survives the producer who created it - when an information goes out, its lifeline becomes independent from the lifeline of the producer. Indeed the decoupling of information and producer and the use of cache allow every nodes which got the content to use it and pass it to every consumer who will issues an interest about it, no matter where the original producer of the data is. 

As a proof of concept, we ran a session of experiments that recreated a situation where a consumer asks for a content after the only producer is gone: there is only one producer running, a consumer asks for a content and when it get it, the producer is switched off. After that, another consumer issues the same interest, that of course can be satisfied only by a CS of some cars around, since there is no more producer for that content.

As expected, the second consumer was able to get the desired content even if the only producer was already gone. Furthermore, thanks to the broadcast nature of the wireless communication that easily allows the spreading of a content, the satisfaction time doesn't seem to be negatively affected by the absence of the producer.

\section{V-NDN at scale}
\begin{figure}[htb]
\begin{minipage}[htb]{\columnwidth}
\centering{
\includegraphics[width=0.9\columnwidth]{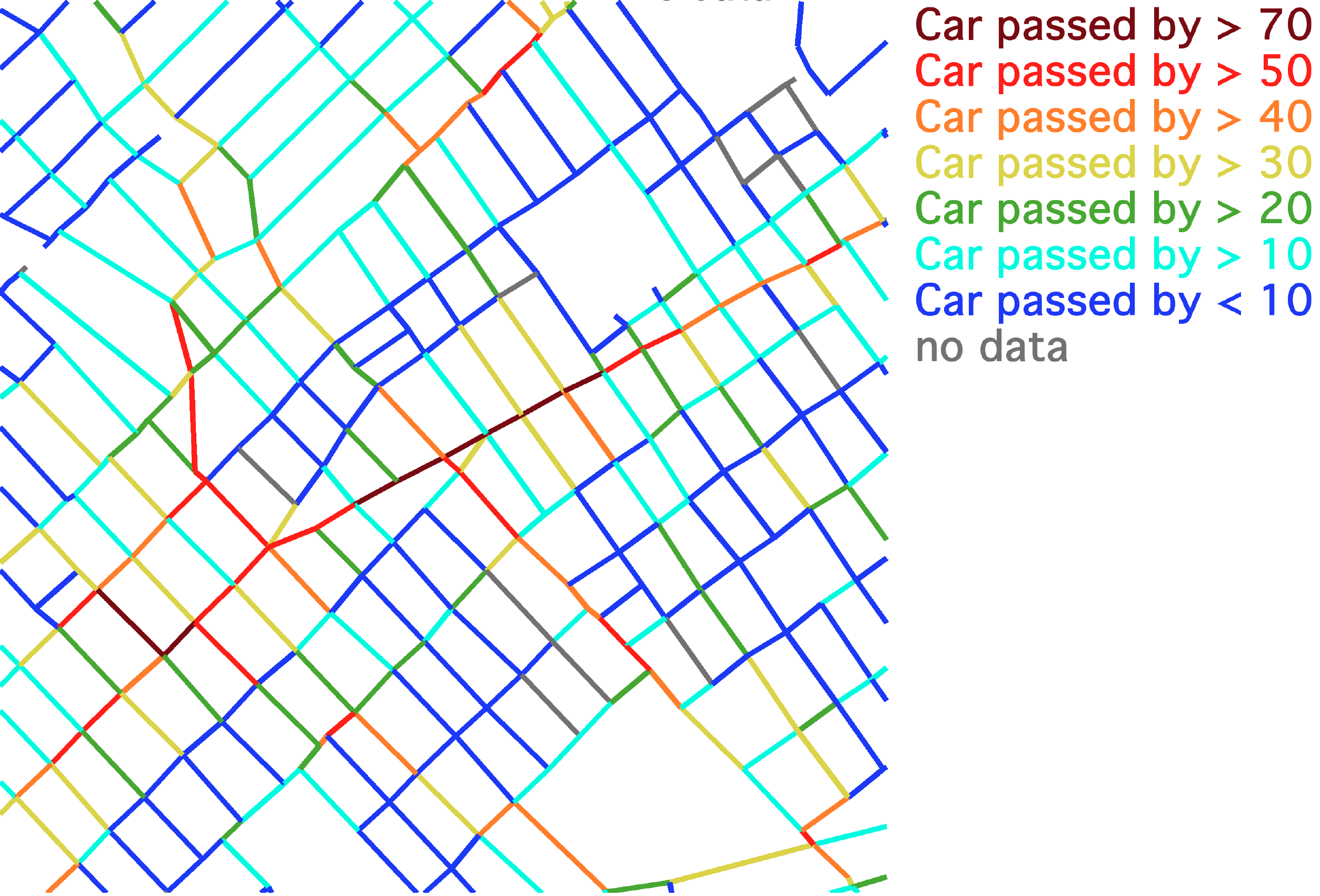}
\caption{\name simulations: Density of vehicles per road segment. }
\label{fig:sim-car-density.pdf}
}
\end{minipage}
\end{figure}

\name scalability has been explored through simulations as it is extremely difficult to run  experiments with hundreds or thousand vehicles. We implemented \name on ndnSIM, the Name Data Network model for NS3~\cite{ndnSIM}. We ran 300 seconds simulation consisting of 695 cars moving in a residential area of 2100 meters per 2100 meters of the city of Los Angeles, CA (34.040569,-118.463308). The cars mobility has been generated using SUMO~\cite{SUMO2012}. To be as much realistic as possible, in the process of generating the mobility, the traffic has been shaped consistently to the importance and size of the roads, as shown in figure~\ref{fig:sim-car-density.pdf}. The radio propagation has been modeled with CORNER~\cite{giordano2010corner} an high fidelity propagation model for urban scenarios that accounts for the presence of buildings as well as fast fading effects. 
All the cars are equipped with a WiFi Ad-Hoc network interface, but only in a subset of them an application (consumer or producer) is running, while the others can be considered as mules.  We simulated the traffic application, running with a fixed number of producers (14\%) while we changed the number of consumer.  The map features 428 road-segments\footnote{In our graph representation of the roadmap segments are defined as  the edges in between intersections (vertex)}. In the area considered there are three road-segment classes: \emph{(i)}  50 6-lanes segments (11.7\%) with a speed limit of 45Mph; \emph{(ii)}  34  4-lanes segments (7.9\%) with a speed limit of 35Mph; and  \emph{(iii)}  344  2-lanes segments (80.4\%) with a speed limit of 25Mph.
\subsection{Results}
Although the packet forwarding strategies for V-NDN are still in their infancy and will certainly improve in the future, the preliminary simulation  results presented in this work serve as a guide to understand \name feasibility at scale even when a highly suboptimal controlled flooding strategy is used.

\begin{figure}[htb]
\begin{minipage}[htb]{\columnwidth}
\centering{
\includegraphics[width=1\textwidth]{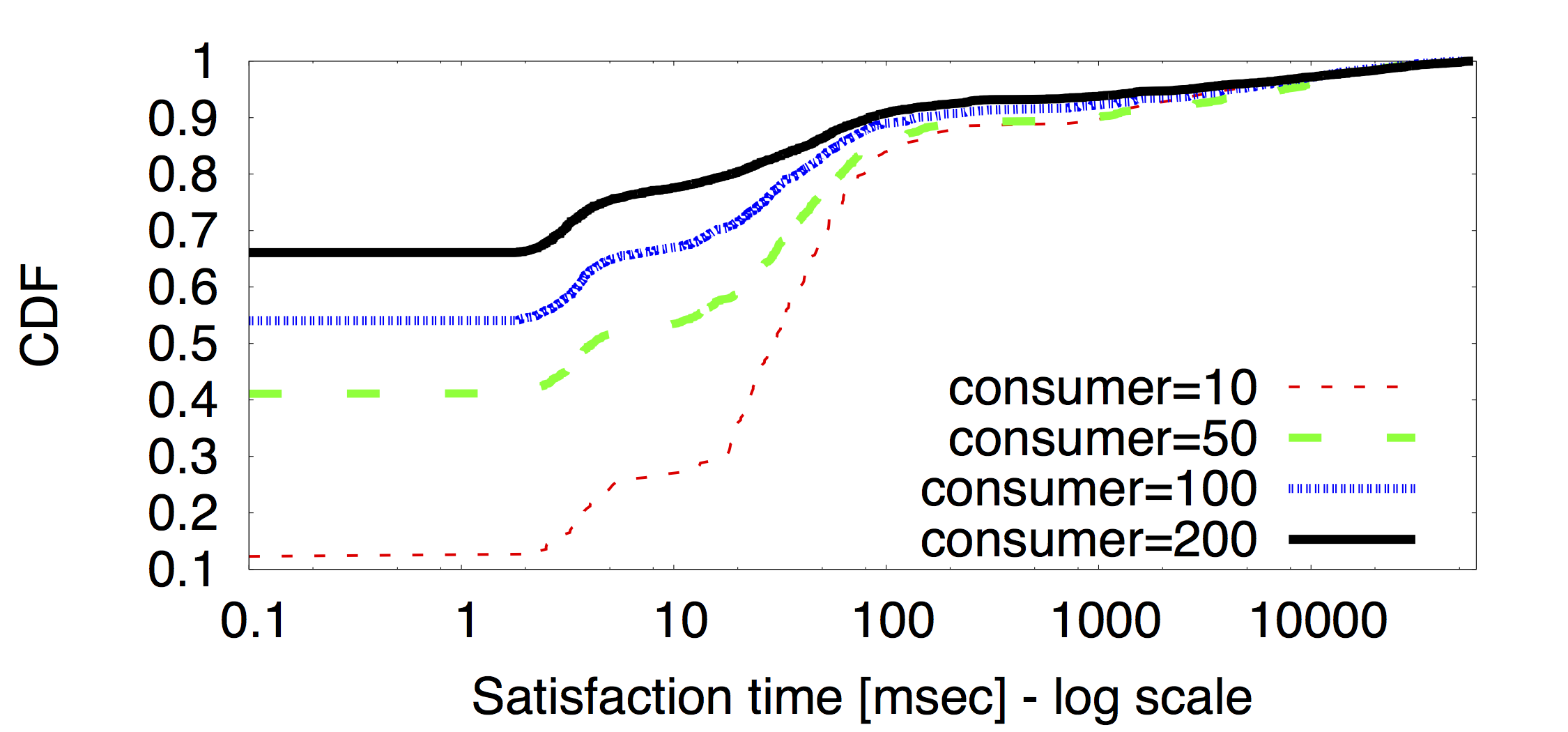}
\caption{\name traffic interest satisfaction time. }
\label{fig:simulation-satisfaction-time.png}
}
\end{minipage}
\end{figure}

\begin{figure}[htb]
\begin{minipage}[htb]{\columnwidth}
\centering{
\includegraphics[width=0.9\textwidth]{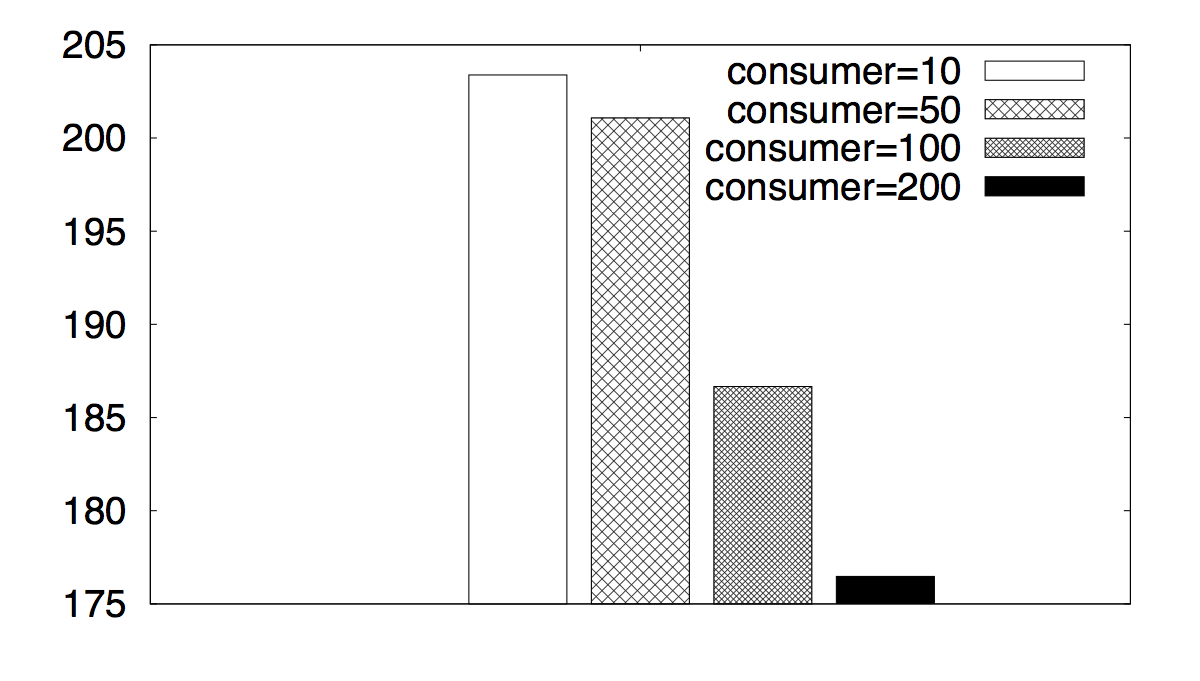}
\caption{\name overhead: number of interest sent in the air for each interest issued by an application. }
\label{fig:overhead}
}
\end{minipage}
\end{figure}

Figure~\ref{fig:overhead} shows how many interest are sent on the network every time the application consumer issues an interest while figure~\ref{fig:simulation-satisfaction-time.png}  shows how much time occurs for a consumer to get the desired content. Both graphs reveal an important factor:  when the number of nodes interested on the same set of information increases, the performance of the entire system in terms of satisfaction time and overhead improves substantially.
Indeed the process of caching overheard information and the NDN fundamental approach of naming content instead of nodes transforms all the mules that get a content into ``producers'' of the content itself:  not only a mule can forward a content that it has just received, collaborating to the forwarding process, but it will also be able in the future to answer to interest for the same content by itself, without the producer intervention, and so without any further propagation of the interest, simply by checking the names. 

\section{Discussion}
Wireless technology advances let us free from stationary communication leading to networking mobile devices. While cellular communications provide a global wireless infrastructure, spectrum scarcity and technology challenges force Mobile Network operator to encourage the use of alternative technologies such as WiFi to keep up with a growing user base and increasing data demands. As matter of fact most of the current cellular providers  in the US and abroad offer only fixed data provision and the once popular all you can eat model is long gone.  

\subsection{Internet untethered}
Multi-homing and mobility enable users to exploit natural information locality an take advantage of opportunistic communication models such as  Ad Hoc communications. Users can retrieve local data reducing the impact on the cellular infrastructure and opportunistically connecting to local hot-spots or other users.  This approach requires a non trivial effort to be performed via IP protocol and comes naturally with \name. 

\subsection{Using names for communication}
V-NDN, by naming data rather than hosts and untying data from IP addresses, can bring substantial benefits to the communication: 
\begin{itemize}
\item It removes the isolation between applications and network transport, allowing forwarding nodes to handle data based on application needs.
\item Communication can start spontaneously, because an infrastructure for the IP addresses assignment is no longer required.
\item Named data facilitate data security, content in the network is signed by the author. 
\item Locally produced data and data with local meaning, as traffic information, are no longer required to be transferred to remote servers;  data that is produced and consumed in loco can remain in loco and be delivered to the consumers by the nodes that are in place, without going through remote servers.
\end{itemize}

\subsection{Remaining Challenges}
Mobile Named Data Networking is still in its early stages. There are several challenges and research problems to be solved to become mainstream as well summarized in~\cite{Tyson2012Survey}. In particular, decrease the communication overhead by limiting the message propagation without affecting the performance of the network is still an open issue. 

Name space is the core of NDN and consequently the name design remains active research. Indeed naming schemes not only concern application, but can positively affects communication performances: use name to help the forwarding process is one of the future step to improve \name architecture.   Beside the name design, efforts will be spent to extend \name not only to vehicle networking, but to all other types of mobile devices including mobile phones.

\bibliographystyle{abbrv}
\bibliography{2013-VNC-Submission}

\end{document}